# Temperature dependence of optical rotation study on parity-violating phase transition of D-, L-, and DL-alanine

Fan Bai [a]   Miao Zhang [a]   Wenqing Wang *[b]   Wei Min [b]   Weimin Du [a]

[a] *Department of Physics, Peking University, Beijing 100871, China*
[b] *Department of Applied Chemistry, College of Chemistry and Molecular Engineering, Peking University, Beijing 100871, China*

**Abstract**– Chiral molecules are characterized by a specific optical rotation angle. An experimental method was presented to dissect the temperature dependence of the optical rotation angle with the molecular chirality of D-alanine, L-alanine and DL-alanine crystals. Salam hypothesis predicted that quantum mechanical cooperative and condensation phenomena may give rise to a second order phase transition below a critical temperature linking the transformation of D-amino acids to L-amino acids due to parity-violating energy difference. The temperature-dependent measurement of the optical rotation angle of D-, L- and DL-alanine crystals provided the direct evidence of the phase transition, but denied the configuration change from D-alanine to L-alanine. New views on Salam hypothesis are presented to demonstrate its importance in the application of low temperature enantiomeric separation and the origin of biochirality.

## 1. Introduction

Two recently published papers by Bonner[1] and Keszthelyi[2] have discussed the extraterrestrial origin of the homochirality of biomolecules and the amplification of tiny enantiomeric excess. Attention is called upon the mechanism for production of extraterrestrial handedness based on Salam phase transition. Salam[3, 4] postulated that chirality among the twenty amino acids, which make up the proteins, may be a consequence of a phase transition which is analogous to that due to BCS superconductivity. Salam suggested a unique mechanism whereby the parity violating energy difference (PVED) between enantiomers might lead directly to a homochiral product. He surmised that there should exist an abrupt phase transition at a critical temperature $T_c$, causing a change in the chirality from D-amino acid to that of more stable "natural" enantiomer, the L-amino acid. Salam proposed the following ways to test his hypothesis:

The first way to test for evidence of the hypothesis (regarding the existence of a phase transition) is to lower the temperature while measuring the optical activity when polarized light is shone upon a particular amino acid crystal. If the polarization vector gets rotated, one may be sure that the appropriate phase transition has taken place. The second way of detecting the process may be by

- Corresponding author. E-mail: wangwq@sun.ihep.ac.cn

measuring differences of specific heats and looking for anomalies in the curve $C = \gamma T + \beta T^3 +\ldots$ like Mizutani et al.[5] have done for the non-amino acids eg. melanins and tumor melanosomes. Note that ideally Salam would be able to compute the values of $T_c$ when electroweak interaction theory is fully worked out. The analogy of the "superfluidity" exhibited by amino acids is to "superfluidity" in superconductors. In the case of superconductivity, the third way for a given amino acid is to apply an external magnetic field and look for the Meissner effect to determine $T_c$.

Figureau *et al.*[6,7] conducted a series of experiments to test the validity of Salam prediction. They observed no changes of optical rotation in solution after exposing both racemic DL-cystine and L-cystine to temperatures ranging from 77K to 0.6K for three and four days, thus reported failing to validate PVED-induced phase transitions predicted by Salam. We argue that Figureau samples were taken out of the cryostat after the cooling period, then heated and dissolved in HCl solution at room temperature for measurement. Therefore, their observing no optical rotation in solution only proves that Salam phase transition is not an irreversible process, but it is not a piece of negative evidence while the transition is reversible. Moreover, their work only denied the configuration change from D-cystine to L-cystine. Based upon our previous studies,[8-10] Salam hypothesis on the phase transition may play as an amplification mechanism instead of a direct configuration change, so the temperature-dependent on-line measurement of optical rotation angle is very necessary for the crystalline DL-alanine, D-alanine and L-alanine.

The rotation of plane–polarized light by chiral molecules arises from the difference of refraction index for left and right circularly polarized light. It is probed at a single frequency in polarimetry measurements of the rotation angle. The angle depends on the chemical bonding and molecular conformation in a molecule. Here we present the experiments to determine the temperature-dependent optical rotation of crystalline D-alanine, L-alanine and DL-alanine.

## Experimental

### 2.1 Characterization of samples

The powder of D-alanine, L-alanine and DL-alanine were obtained from Sigma Chemical Co. The amino acids single crystals were well-formed crystals elongated along the *c* axis, which were grown by slow evaporation of saturated aqueous solutions at 4°C, then washed with absolute alcohol, evacuated and kept in a desiccator. The characterization of D- and L-alanine crystals was performed by the element analysis (C, H and N) and X-ray diffraction analysis. It indicates that D-alanine and L-alanine are pure single crystals containing no crystalline water. The crystal structures are orthorhombic with the same space group $P2_12_12_1$, Z=4, with lattice constants a = 0.60388(1), 0.60344(5) nm, b = 1.23670(3), 1.23668(8) nm, c = 0.58000(2), 0.57879(3) nm of D-and L-alanine (300K), $\alpha = \beta = \gamma = 90°$, which agrees with Simpson data[11], a =0.6032, b=1.2343, c=0.5784nm (298K). The rotation angle $\zeta$ of the D- and L- alanine solution was measured on Polarimeter PE-241MC with the wavelength of 589.6 nm at 293K. By using the formula of $[\alpha] = \zeta / (L+C)$, the corresponding $\alpha$ values of D- and L-alanine were shown to be the same absolute value and of opposite signs. The above data of D- and L-alanine confirm that the experiment is starting from identical material of D- and L-alanine without contaminants.[10]

## 2.2 Temperature-dependent optical rotation angle measurement

D-, L- and DL-amino acids are characterized by a specific optical rotation angle, through which plane-polarized light is rotated on when passing through an enantiomerically enriched solution. The accuracy of the most sophisticated polarimeter is $10^{-4} \sim 10^{-5}$ degree, but it is used to measure the optical rotation angle in aqueous solution at room temperature. In order to study on Salam Phase Transition ($T_c$ was predicted as about 250K), we have to measure the optical rotation in crystalline amino acid while the temperature is much lower than the freezing point of aqueous solution. The variation of optical rotation angle of crystal with temperature is complicated causing by subtle changes in molecular structure e.g. changes of atomic coordinates and ability of certain groups (such as carboxylate, methyl and amino group) to rotate.[12] The angle depends on chemical bonding and molecular conformation as yet poorly understood way. Linking the rotation angle to molecular structure is a challenge of fundamental and practical importance. Here we demonstrate an experimental method to determine the temperature- dependent variation of optical rotation angle of crystalline alanine enantiomers.

Linearly polarized light of wavelength 632.8 nm from He-Ne laser source passes through a transparent chiral material of length $d$ that is located in vacuum temperature-control system. The direct measurement of the polarization direction of a light beam e.g. the rotation signal must be converted into an amplitude signal with the aid of a polarization-optical analyzer. Wollaston prism is a beam splitter that splits the linearly polarized light beam into two component beams polarized normal to each other. Because the maximum sensitivity and linearity of measurement are obtained by adjusting an angle of 45° between the analyzer and the initial plane of polarization, the component beam intensities $J_1$ and $J_2$ are measured with two highly accurate photodetectors by adjusting the transmitting direction of the Wollaston prism to be ± 45°(Figs.1-2)

$$J_1 = J_0 \cos^2 (45°-\varphi) = J_0 \ (1 + \sin 2\varphi)/2 \qquad (1)$$
$$J_2 = J_0 \cos^2 (45°+\varphi) = J_0 \ (1 - \sin 2\varphi)/2 \qquad (2)$$
$$J_1 - J_2 / J_1 + J_2 = \sin 2\varphi \qquad (3)$$
$$\varphi = 1/2 \ \mathrm{arc} \sin (J_1 - J_2 / J_1 + J_2) \qquad (4)$$

Here $J_0$ is the intensity in front of the analyzer, $J_1$ and $J_2$ is the light intensity behind the analyzer, respectively.[13]

The temperature-dependent optical rotation angles of cryatalline DL-alanine, D-alanine and L-alanine have been determined from 230K to 290K. The crystals were pre-located in the temperature-control vacuum sets, and the beam intensity $J_1$ and $J_2$ was measured by two independent highly accurate photodetectors. The optical rotation angle $\varphi$ (degree) was calculated by equation (1) to (4) and the data were shown in Tables 1-3.

The optical rotation angle $\varphi$ is defined as
$$\varphi_L = 2\pi n_L d / \lambda \qquad (5)$$
$$\varphi_R = 2\pi n_R d / \lambda \qquad (6)$$
$$\varphi = 1/2 (\varphi_L - \varphi_R) = \pi (n_L - n_R) d / \lambda \qquad (7)$$

Here $n_L$ and $n_R$ are the refraction index of left circularly polarization light and right polarization light, *d* is the thickness of measuring crystals, $\lambda$ is the wavelength in vacuum (Fig. 3).

The temperature-dependent optical rotation angle of DL-alanine was shown in Fig.4. The $\varphi$ value was equal to 0.50° ± 0.10° (approach to zero) from 288K to 270K. It proves that DL-alanine is truly racemic. When the temperature continuously decreased from 260K through 250K to 234K, the $\varphi$ value was rapidly increased to 4.26° shown an obvious characteristic variation. According to the Salam prediction, we may be sure that a second-order phase transition has taken place around 261± 1 K.   The temperature-dependent optical rotation angle of crystalline L-alanine was shown in Fig.5. The $\varphi$ value was fluctuated in the range of 9.20°~ 10.01° from 278K to 233K and shown maximum 10.01° at 260 ± 1 K. It coincides a phase transition of lattice mode happened in L-alanine producing a small peak of optical rotation angle.[10, 14] In contrast, the $\varphi$ value of D-alanine was obviously increased from -9.09° (284K) at a maximum –3.98°(263K) then decreased to −10.37°(233K) appeared a parabolic reversible transformation (Fig.6). It demonstrated that a crucial form of the transition temperature $T_c$ involved dynamical symmetry breaking. The structures of DL-alanine, D-alanine and L-alanine belong to the orthorhombic system, but the space group is $P2_12_12_1$ for the D- (or L-) alanine and $Pna2_1$ for the DL-alanine[15], respectively. It is reasonable that the optical rotation angle of DL-alanine $\varphi_{DL}$ is not apparently equal to the sum of D-alanine $\varphi_D$ plus L-alanine $\varphi_L$ from the Tables 1-3.

## 3.  Results and Discussion

**3.1 Parity violation in molecular alanine system**
For the aim to understand the parity violation energy ($E_{PV}$) in chiral molecules, in terms of an approximate Hamiltonian, there are relativistic and nonrelavistic two main approaches to $E_{PV}$ computations. From a nonrelativistic wave function, $E_{PV}$ is estimated as the second–order perturbative energy due to the coupling of spin-orbit (SO) and parity-violation (PV) Hamiltonians. These are given by the expressions[16]

$$H_{SO} = \frac{\beta^2}{\hbar} \sum_{i,N} Z_N |\vec{r}_i - \vec{R}_N|^{-3} \vec{\sigma}_i \cdot (\vec{r}_i - \vec{R}_N) \times \vec{p}_i \tag{8}$$

$$H_{PV} = -\frac{G_F}{4\sqrt{2} m_e c} \sum_{i,N} Q_{W,N} \{\vec{p}_i \cdot \vec{\sigma}_i, \delta(\vec{r}_i - \vec{R}_N)\}_+ \tag{9}$$

where $\beta$ indicates Bohr magneton, $Z_N$ nuclear charge, $m_e$ electron mass, *c* the speed of light, **r** and **p** electronic position and momentum, **σ** the vector of Pauli spin matrices, and $\mathbf{R}_N$ is a nuclear coordinate. The explicit form for $E_{PV}$ is

$$E_{PV} = -\frac{2}{\hbar} \sum_{j \neq a} \frac{\text{Re}(\langle a|H_{PV}|j\rangle\langle j|H_{SO}|a\rangle)}{\omega_{ja}} \tag{10}$$

here a and j refer to ground and excited state wave function, respectively, and $\hbar\omega_{ja}$ is an energy difference.

$E_{PV}$ depends on the weak charge $Q_{W,N}$ of the nuclei and hence on the atomic number $Z_N$. This dependence has been estimated theoretically for atoms as $Z_N^3$ and for molecules as $Z_N^5$. It is sufficient to note that heavy nuclei contribute more to $E_{PV}$ than light ones. In the case of alanine, without heavy atoms contribute to $E_{PV}$, but in most orbits the atomic electron is virtually beyond reach of the weak force if the orbit is highly eccentric or elongated, such as the single electron in the α-hydrogen atom, the electron can come close enough to the nucleus for the weak force to exert a more powerful influence. We emphasize to consider spin-orbit (SO) term. For each electron and each nucleus, the SO term lowers the total energy when the angular momentum of the electron about the nucleus is antiparallel to the electronic spin. The nonrelativistic PV Hamiltonian gives a negative contribution to the energy when the canonical momentum **p** of an electron on a nucleus is aligned with the electron spin. By estimating the direction of the SO induced momentum for each possible spin orientation, it is possible to predict the sign of $E_{PV}$ for a given molecule. Considering how the induced momentum changes with the geometry we can estimate which configurations are likely to have higher values of $E_{PV}$.

The ordinary parity-conserving electromagnetic forces between the electrons and the nuclei in the molecule, however, tend to align the axis of each electron's orbit against its axis of spin. This phenomenon is referred to as spin-orbit coupling. For a right-handed helical molecule, spin-orbit coupling favors down-spiraling for spin-up electrons and up-spiraling for spin-down electrons. In either case the spin axis of the electron tends to be aligned against the electron's direction of motion. As a result, molecules display regions of differing electron chirality. An important consequence of the weak Z force between electrons and nuclei is that all atoms are chiral (Fig. 7). On a slightly larger scale, the Z force causes a chiral molecule to exist in a higher- or lower-energy state than that of its enantiomer.[17]

In order to investigate the behavior of $E_{PV}$ against simple geometric changes, Faglioni *et al.*[16] denoted that the parity violation energy difference ($E_{PV}$) is related to the geometric distortions and group rotations under temperature lowering. Each stretching and bending of chemical bonds is associated with larger changes in electronic configuration than are generally expected for the case of rotation about σ bonds. $E_{PV}$ depends on the electrons of the spin perpendicular to the $C_\alpha$-H bond. Therefore, $E_{PV}$ is expected to increases when the direction of motion of the electrons moves away from the direction of the $C_\alpha$-H bond. The temperature–dependent $^1$H –MAS ssNMR spectra of D-alanine [18] confirm this behavior. The peak of $^1$H –α C showed an obvious up-shielding for the decrease of the temperature. It was explained that the shielding effect of metal hydrogen atom is higher than the hydrogen of α–carbon in D-alanine molecule. Likewise, for the case of $C_\alpha$-H stretching, we expect $E_{PV}$ to be proportional to the electronic momentum induced by spin-orbit coupling due to hydrogen nucleus measured on the carbon nucleus. This momentum should decrease with the distance between the nuclei. $E_{PV}$ should then decrease with the $C_\alpha$-H stretching and eventually vanish at infinite distance. This behavior is also consistent with the temperature-dependent Raman spectra[19] and optical rotation parabolic reversible transformation of D-alanine. When temperature exceeds the transition temperature $T_c \approx 250K$, the spectra of $C_\alpha$–H modes at 1305cm$^{-1}$ ($C_\alpha$–H bending), 2964 cm$^{-1}$ ($C_\alpha$–H stretching) of D-alanine show a downward shift of 1.7 cm$^{-1}$ variation and relative intensity decreasing, confirmed the existence of phase transition of D-alanine.

**3.2 Vibrationally generated ring currents in alanine enantiomers**

The calculation of optical rotation angle is related to the angular frequency of the incident radiation ($\omega$) and the elements of the electric dipole and magnetic dipole, which can be written down as:

$$\varphi = \frac{8\pi NL}{3\hbar c} \sum_j \frac{\text{Im}\langle 0|\mu|j\rangle \bullet \langle j|m|0\rangle \omega^2}{\left(\omega_{j0}^2 - \omega^2\right)} \tag{11}$$

Here $|0\rangle$, $|j\rangle$ denotes the ground and exited state wave functions in Dirac presentation, respectively, and $\omega_{j0} = \omega_j - \omega_0$ is the associated excitation frequency. Here $\mu$ and m are the electric and magnetic dipole operators. N means the total number of atoms in a unit volume. $L$ denotes the length of light pass, $\hbar$ is Plank constant and $c$ is the light velocity.

This equation shows that the interference of the electric dipole and magnetic dipole produces the optical rotation effect in optical active medium. To perform this calculation on alanine crystal, not only the contributions of single atoms should be included, but also all the chemical bonds and the molecular conformation must be considered. T. B. Freedman [20] proposed a model via vibrationally generated electronic ring currents to explain the optical rotation effect. In the case of alanine crystal, the $C_\alpha$–H stretch generates an oscillating electronic current in a molecular ring, adjacent to the methine bond, which is closed by hydrogen bonding. This oscillating ring current gives rise to a large magnetic dipole transition moment. As depicted in Fig. 8 for the $C_\alpha$–H contraction (or lengthening) in D-alanine, positive (or negative) current flowing in the direction N→$C_\alpha$ when electrons are injected into the ring by the $C_\alpha$–H contraction (or lengthening), produces a magnetic dipole transition moment, $m$, with a component in the direction of the electric dipole transition moment $\mu$. The rotational strength R= Im ($\mu \bullet m$), is positive (or negative), as observed experimentally. The main difference between enantiomers lies in that they produce opposite optical rotation due to the opposite sign of Im ($\mu \bullet m$). The vibrational frequencies of $C_\alpha$–H bending and stretching modes are very sensitive to the change of electric dipole moment $\mu$. Raman spectra study indicates the difference of the electric dipole moment $\mu$ between two enantiomers when temperature is below 250K. This result is of great importance to the understanding of the temperature-dependent optical rotation results.

**3.3 Atomic contribution to the optical rotation angle as a probe of molecular chirality**

Chiral molecules are characterized by a specific rotation angle. Recent developments in methodology allow computation of both the sign and the magnitude of these rotation angles. However, the individual contributions that atoms and functional groups make to the optical rotation angle to the molecular chirality has remained elusive.[12] A more direct way to testify Salam phase transition is to conduct temperature-dependent X-ray diffraction on alanine enantiomers. If there is a configuration change from D-alanine to L-alanine, it will be easy to observe an abrupt change in atom coordinates. The data for D-/L- alanine single crystals are collected at 300K, 270K and 250K on a Rigaku RAXIS-RAPID imaging plate diffractometer with Mo-K$_\alpha$ ($\lambda$ =0.71069 Å) radiation. It was not found the configuration change from D-alanine to L-alanine at 250K.

However, there is a significant difference between the dihedral angles of L- and D-alanine at 250K (D-alanine 43.97°, L-alanine 45.72°) originating from the atomic coordinate variations in the course of temperature lowering. Dihedral angles was calculated from the atomic coordinates of O(1) O(2) C(1) C(2) H(4) of D- and L-alanine. Theoretical physicists use symmetry to predict a minimal nodal pattern of $E_{PV}$ as a function of the dihedral angle φ. Berger and Quack's study [21,22] in a detailed analysis of dynamic chirality proved that the dihedral angle between the $O_2C$ and $C_α$-H planes plays an important role in determining the intrinsic energies of the alanine molecules. This difference has been used in the calculation of parity violating energy difference ($E_{PV}$). According to Quack theoretical method by means of highest level ab initio studies (MC-LR), the value is $1.2 \times 10^{-19}$ Hartree ($3.3 \times 10^{-18}$ eV/molecule), namely L-alanine is more stable than D-alanine. [23]

From Freedman's model, we can make a sound explanation of the phenomenon observed in our experiments. The existence of the phase transition suggested by Salam is verified by our experiments. NMR and Raman spectra have confirmed that the phase transition can induce a change in the electric dipole moment $\mu$. In addition, the slight variation of atoms' position and a change of dihedral angle (45.52°→ 43.97°)[19], which will bring the change in the magnetic dipole moment $m$. Since $\mu$ and $m$ all undergo a change in their magnitude, it is natural to observe the variation of rotation angle in this phase transition because it is closely related to Im ($\mu \bullet m$). The variation of rotation angle cannot be interpreted as the configuration change of D→L. As we have already presented, the optical rotation is closely related to Im ($\mu \bullet m$), so changes of magnitude of Im ($\mu \bullet m$) will also bring about this phenomenon.

### 3.3 Conclusion

Our series of experiments confirm the existence of a phase transition in alanine. D- and L-alanine experience different behavior below $T_c \approx 250K$, which may be important in the application of enantiomeric separation at low temperature. The parity violation effect of weak interaction has made enantiomers asymmetry. As well known in all the molecules examined, the estimated $E_{PV}$ is several orders of magnitude smaller than current experimental resolution. However, as we have illustrated, D-, L-, and DL- alanine crystals display obvious different behavior in the phase transition process. A possible explanation could be: the minor difference between D- and L-enatiomers (PVED: about $10^{-18}$ - $10^{-17}$ eV) has been enlarged to a detective level during this phase transition due to quantum mechanical cooperative and condensation phenomena.

A kinetic model has been suggested by Kondepudi and Nelson (KN)[24, 25] supposing the energy difference of $\approx 10^{-20}$ hartree, which makes natural enantiomers thermodynamically more stable. KN showed that symmetry can be broken via a catastrophic reaction mechanism, e.g., reaction volume of $4 \times 10^9$ liter (i.e., a lake of 1 km × 1km × 4m), concentrations of order $10^{-3}$ M, and reaction rates of $10^{-10}$ M sec$^{-1}$. In these conditions $10^4$ yrs would be necessary for an excess of 98% of the favored enantiomer to develop from a primordial racemic mixture. In any event, an energy difference of $\approx 10^{-20}$ hartree is too small to make the KN model really effective which has prompted other scientist to wonder whether the electroweak force could actually play a fundamental role in the appearance of chirality in terrestrial biochemistry[26]. But if the tunneling mechanisms have caused second-order phase transition between enantiomers below a critical temperature, as suggested by Salam relying on the WNC hypothesis, the energy difference may be enlarged many times before the nonlinear chemical reaction. This amplified energy difference, as we have discovered at the critical phase transition temperature, will become the foundation of nonlinear amplification mechanism.

In summary, we emphasize the significance of Salam phase transition in the evolution of homochirality instead of an ultimate solution to the problem, as it may actually play as the first step of amplification mechanism. It connects the microcosmic difference $E_{PV}$ between biomolecular enantiomers with nonlinear process in a macrocosmic biological system. It solves the long-term debating suspicion that $E_{PV}$ is too minor to be enlarged directly by nonlinear process. Combining the existence of $E_{PV}$, Salam phase transition and nonlinear amplification mechanism, we may propose a sound way to understand the chemical evolution of homochirality as depicted in Fig.9.

## Acknowledgments

This research was supported by the grant of 863 program (863-103-13-06-01) and by the grant of National Natural Science Foundation of China (29672003).

## References

1. Bonner, W.A. *Chirality.***2000,** 12**,**114-126.
2. Keszthelyi, L. *Orig. Life. Evol. Biosphere*. **2001,** 31**,** 249-256.
3. Salam, A. *Phys Lett B*. **1992,**288,153-160
4. Salam, A. *J.Mol.Evol*. **1991**,33,105-113
5. Mizutani, U.; Massalski,T.B.; McGiness, J.E.; Corry, P. M. *Nature*. **1976,** 259**,** 505-507.
6. Figureau, A.; Duval, E.; Boukenter, A. Chemical Evolution: Origin of Life. Edited by Cyril Ponnamperuma and Julian Chela-Flores. A Deepak Publishing, Hampton Virginia, USA. **1993,**157-164
7. Figureau, A.; Duval, E.; Boukenter, A. *Orig. Life. Evol. Biosphere*.**1995,** 25,211-217
8. Wang, W.Q.; Shen, X.R.; Yang, H. S.; Zhuang, Z.Z.; Lou, F. M.; Chen, Z. J. *J.Biol. Phys*. **1994,** 22, 65-71
9. Wang, W.Q.; Shen, X..R.; Jin, H. F.; Wu, J. L.; Yin, B.; Li, J. W.; Zhao, Z. X.; Yang, H. S.; Lou, F. M.; Zhuang, Z.Z.; Yu, G. Y.; Shi L.; Chen, Z. J. *J.Biol.Phys*. **1996,** 20,247-252.
10. Wang, W.Q.; Yi, F.; Ni Y. M. Zhao, Z. X.; Jin, X. L.; Tang, Y. Q. *J.Biol.Phys*. **2000,** 26,51-65.
11. Simpson, H.J.; Marsh, R.E.  *Acta Cryst*. **1996**,20,550-555.
12. Kondru, R. K.; Wipf, P.; Beratan, D. N. *Science* **1998***,* 282 (18) 2247-2250.
13. Pepp, A.; Harms, H. Magnetooptical current transformer 1: Principles. *Appl. Optics*. **1980,** 22, 3729-3734.
14. Wang, W. Q.; Liang Z. *Acta Phys.-Chim. Sin*. **2001,** 17(2), 1077-1085.
15. Nandhini, M. S.; Krishnakumar R. V.; Natarajan, S.  *Acta Cryst*. **2001**,57**,** 614-615.
16. Faglioni, F.; Lazzeretti, P. *Physical Review E***,**  **2002,** 65, 011904-1-11.

17. Hegstrom, . A.; Kondepudi, D.K. *Scientific American* **1990**, 1, 98-105.
18. Wang W. Q.; Min, W.; Liang, Z.; Wang, L. Y.; Chen, L.; Deng, F. *Biophysical Chemistry,* **2002,** accepted.
19. Wang, W. Q.; Min, W.; Bai, F.; Sun, L.; Yi, F.; Wang, Z. M.; Yan, C. H.; Ni, Y. M.; Zhao, Z. X. *Tetrahedron Asymmetry* **2002**, 13, 2427-2432.
20. Freedman,T. B.; Balukjian, G.A.; Nafie, L. A. *J. Am..Chem.Soc.* **1985,**107**,** 6213-6222.
21. Berger, R.; Quack, M. *Chemphyschem.* **2000,**1,57-60.
22. Quack, M.; Stohner, J. *Phys. Rev. Lett.* **2000**, 84, 3807-3810.
23. Wang, W.Q.; Sun, L.; Min, W.; Wang, Z. M. *Acta Phys.-Chim. Sin.* **2002,** 18(10), 871-877.
24. Kondepudi, D. K.; Nelson, G. W. *Phys. Rev. Lett.* **1983,** 50, 1023- 1026. .
25. Kondepudi, D. K.; Nelson, G. W. *Nature.* **1985,** 314. 438- 441.
26. Zanasi, R.; Lazzeretti, P.; Ligabue, A.; Soncini, A. *Phys. Rev. E* **1999**, 59(3), 3382-3385.

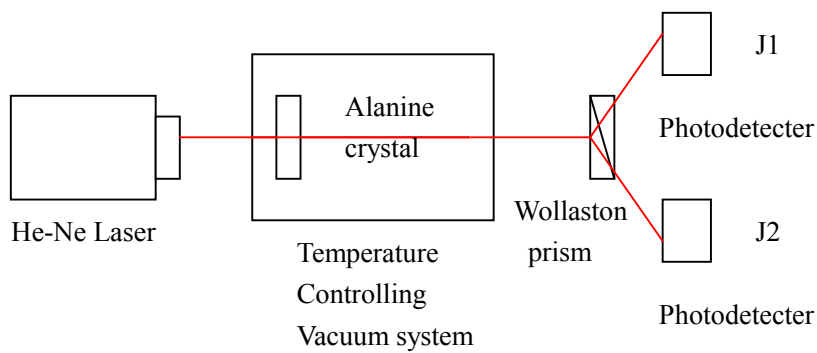

**Fig.1 Schematic diagram of the apparatus**

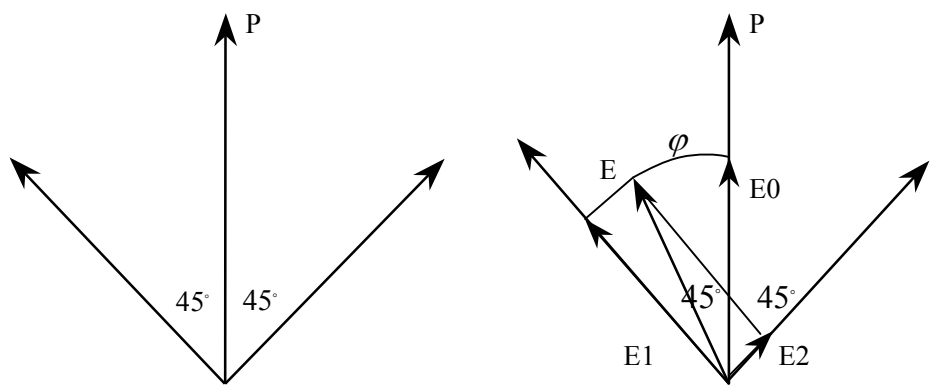

**Fig.2  Resolution of polarization vector through Wollaston prism**

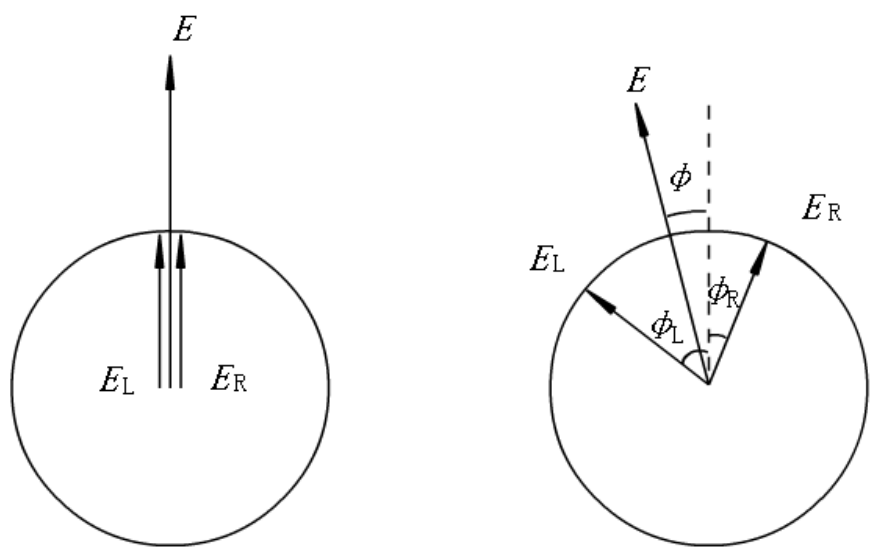

**Fig. 3 Illustration of optical rotation in chiral material**

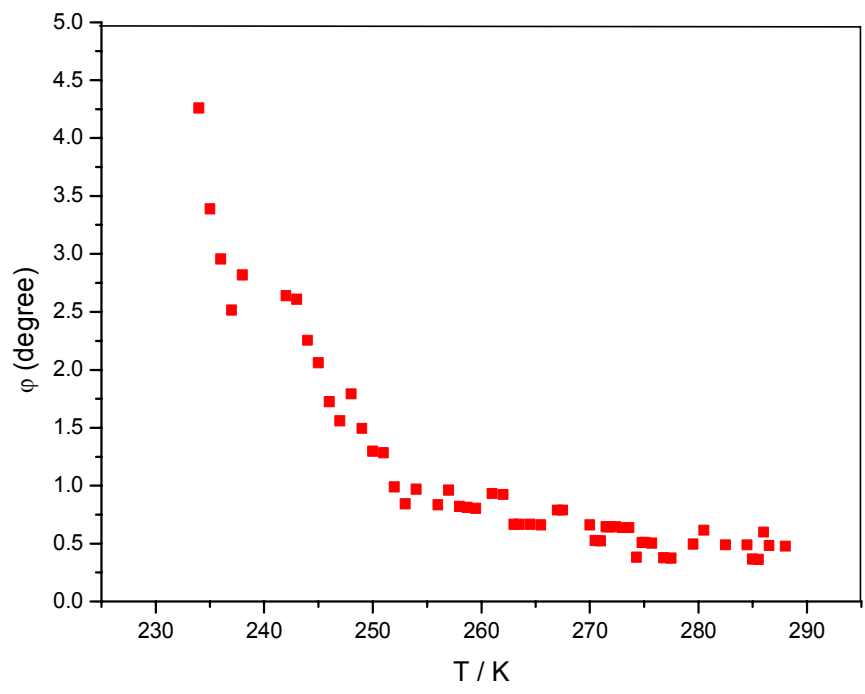

**Fig. 4 Temperature-dependent rotation angle of DL-alanine**

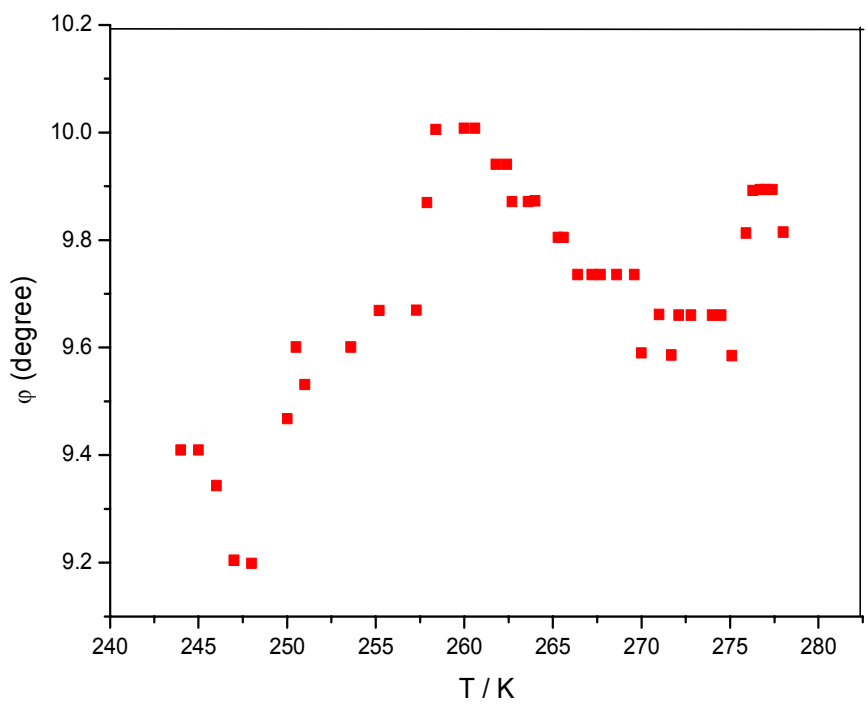

**Fig.5 Temperature-dependent rotation angle of L-alanine**

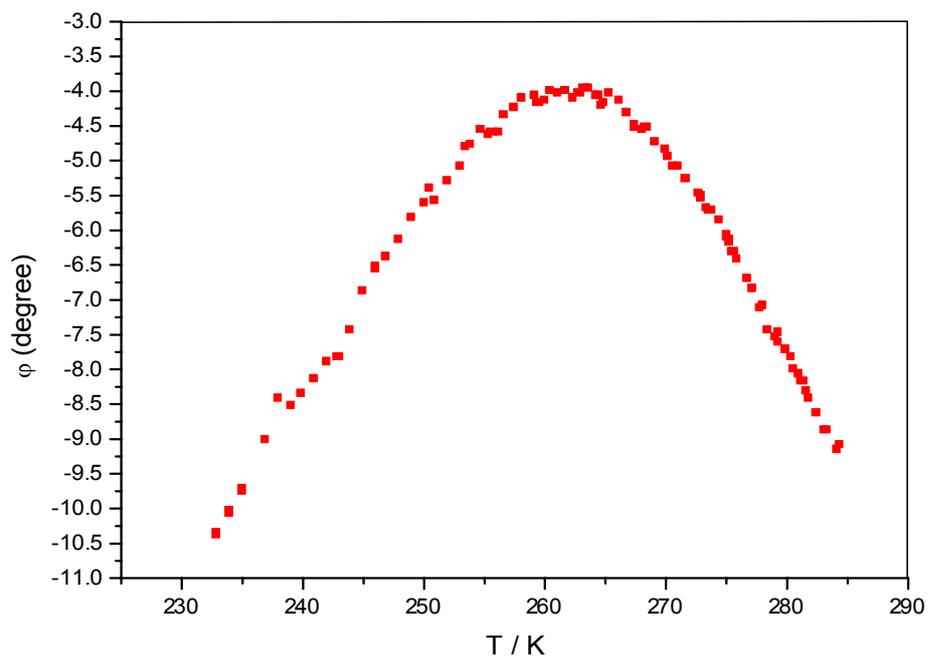

**Fig. 6 Temperature-dependent rotation angle of D-alanine**

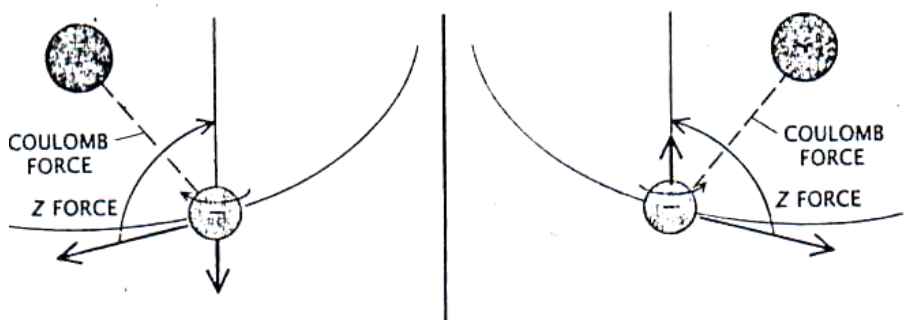

**Fig.7 Atoms become chiral under Z force[17]**

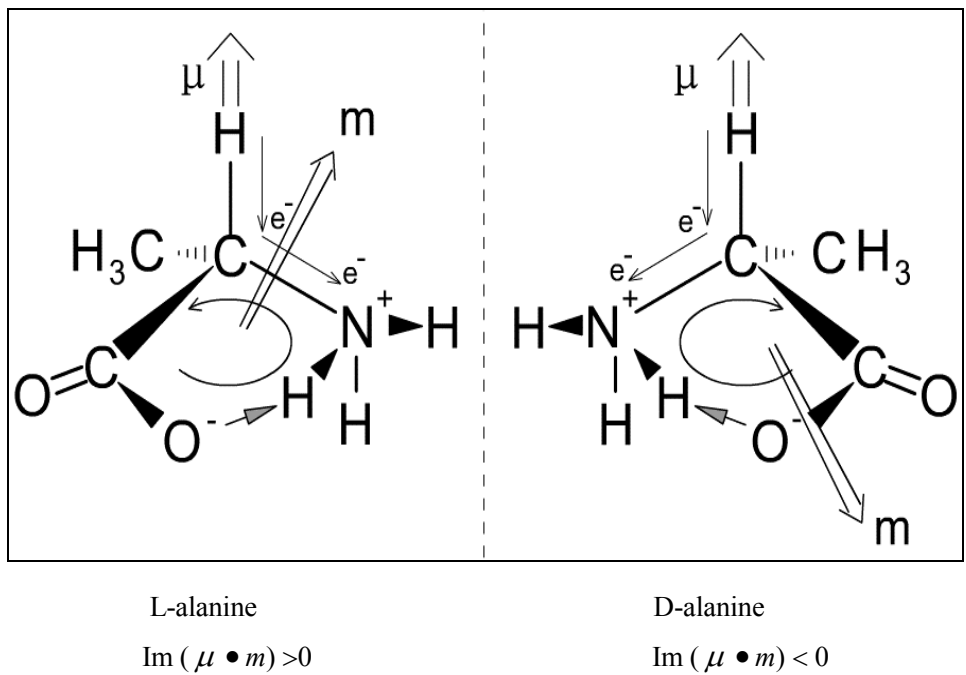

        L-alanine                        D-alanine

       Im $(\mu \bullet m) > 0$                Im $(\mu \bullet m) < 0$

**Fig.8  Vibrationally generated ring currents in alanine enantiomers**

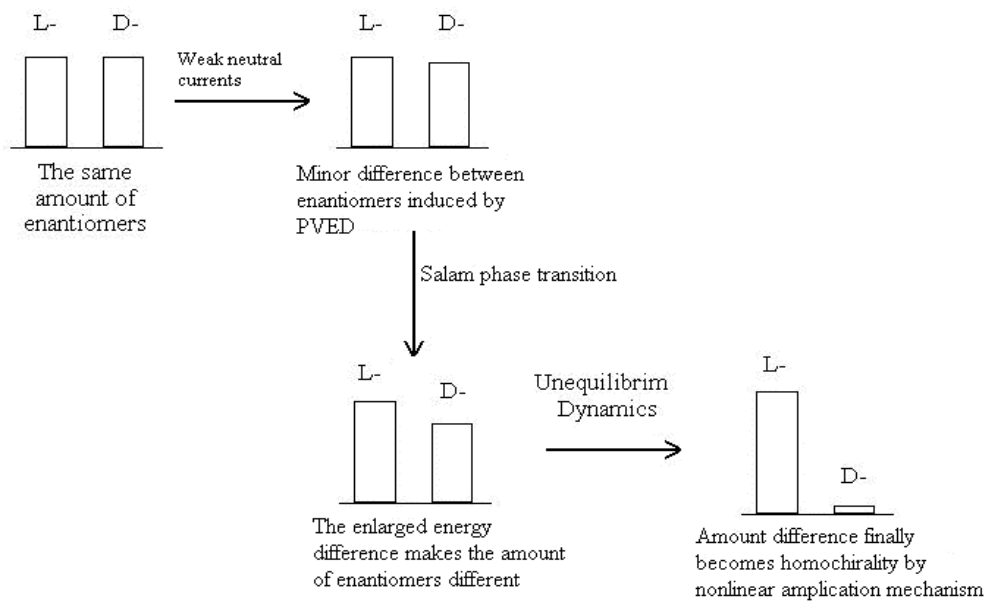

**Fig. 9 A possible evolution process of homochirality**

13